\documentclass[conference,romanappendices,10pt]{IEEEtran}

\IEEEoverridecommandlockouts

\usepackage{algorithm,algorithmic,amsmath,amssymb,amsthm,bbm,caption,cite,color,comment,graphicx,microtype,multicol,setspace,subcaption,tabularx,url}
\usepackage[USenglish]{babel}
\usepackage[T1]{fontenc}
\usepackage[utf8]{inputenc}

\usepackage{enumitem}
\setitemize{itemsep=0mm,topsep=0mm,parsep=0pt,partopsep=0pt}

\makeatletter
\newcommand\subparagraph{%
  \@startsection{subparagraph}{5}
  {\parindent}
  {3.25ex \@plus 1ex \@minus .2ex}
  {-1em}
  {\normalfont\normalsize\bfseries}}
\makeatother
\usepackage{titlesec}
\let\subparagraph\relax

\usepackage{titlesec}
\let\subparagraph\relax
\titlespacing{\section}{0pt}{5pt plus 2pt minus 1pt}{3pt plus 1pt minus 0pt}
\titlespacing{\subsection}{0pt}{4pt plus 2pt minus 1pt}{2pt plus 1pt minus 0pt}
\usepackage[figurename=Fig.,font=small]{caption}

{\newtheorem{theorem}{Theorem}}
{}
{}
{}
{}
{}
{\newtheorem{corollary}{Corollary}}

\renewcommand{\a}{\mathbf{a}}

\renewcommand{\d}{\mathbf{d}}

\newcommand{\p}{\mathbf{p}}

\renewcommand{\r}{\mathbf{r}}

\newcommand{\x}{\mathbf{x}}
\newcommand{\y}{\mathbf{y}}
\newcommand{\z}{\mathbf{z}}

\newcommand{\A}{\mathbf{A}}

\newcommand{\C}{\mathbf{C}}

\renewcommand{\H}{\mathbf{H}}
\newcommand{\I}{\mathbf{I}}

\renewcommand{\P}{\mathbf{P}}

\newcommand{\R}{\mathbf{R}}

\newcommand{\V}{\mathbf{V}}

\newcommand{\Y}{\mathbf{Y}}
\newcommand{\Z}{\mathbf{Z}}

\newcommand{\setC}{\mathcal{C}}

\newcommand{\setN}{\mathcal{N}}

\newcommand{\setQ}{\mathcal{Q}}
\newcommand{\setR}{\mathcal{R}}
\newcommand{\setS}{\mathcal{S}}

\newcommand{\Compl}{\mbox{$\mathbb{C}$}}

\newcommand{\Exp}{\mathbb{E}}
\newcommand{\herm}{\mathrm{H}}
\renewcommand{\Im}{\mathrm{Im}}

\renewcommand{\Re}{\mathrm{Re}}
\newcommand{\sgn}{\mathrm{sgn}}

\newcommand{\tran}{\mathrm{T}}
\newcommand{\Var}{\mathbb{V}}

\newcommand{\El}{\mathsf{E}_{\ell}}
\newcommand{\rmp}{\textrm{p}}
\newcommand{\Vl}{\mathsf{V}_{\ell}}

\usepackage{fancyhdr}
\fancypagestyle{firstpage}{
  
  \fancyhead[R]{{\scriptsize 1}}
  \fancyfoot[C]{{\footnotesize \begin{singlespace} \textcopyright 2021 IEEE. Personal use of this material is permitted. Permission from IEEE must be obtained for all other uses, in any current or future media, including reprinting/republishing this material for advertising or promotional purposes, creating new collective works, for resale or redistribution to servers or lists, or reuse of any copyrighted component of this work in other works. \end{singlespace}}}}

\title{Uplink Data Detection Analysis \\ of 1-Bit Quantized Massive MIMO \vspace{-1mm}}
\author{Italo Atzeni and Antti Tölli \\
Centre for Wireless Communications, University of Oulu, Finland \\
Emails: \{italo.atzeni, antti.tolli\}@oulu.fi
\thanks{\vspace{-3mm}

The work of I.~Atzeni was supported by the Marie Sk\l{}odowska-Curie Actions (MSCA-IF 897938 DELIGHT). The work of A.~Tölli was supported by the Academy of Finland under grant no. 318927 (6Genesis Flagship).} \vspace{-2mm}}

\begin{document}

\maketitle

\thispagestyle{firstpage}

\begin{abstract}
This paper presents an analytical framework for the data detection in massive multiple-input multiple-output uplink systems with 1-bit analog-to-digital converters (ADCs). Considering the single-user case, we provide closed-form expressions of the expected value and the variance of the estimated symbols when maximum ratio combining is adopted at the base station (BS) along with their asymptotic behavior at high signal-to-noise ratio (SNR). These results are exploited to enhance the performance of maximum likelihood detection by taking into account the dispersion of the estimated symbols about their expected values. The symbol error rate with 1-bit ADCs is evaluated with respect to the number of BS antennas, the SNR, and the pilot length used for the channel estimation. The proposed analysis highlights a fundamental SNR trade-off, according to which operating at the right SNR considerably improves the data detection accuracy.
\end{abstract}

\section{Introduction} \label{sec:INTRO}

\vspace{-0.5mm}

Beyond-5G wireless systems are expected to exploit the large amount of bandwidth available in the mmWave band and raise the operating frequencies up to 1~THz \cite{Raj20}. In this context, fully digital architectures allow to truly capitalize on the massive multiple-input multiple-output (MIMO) arrays to implement highly flexible beamforming and serve more user equipments (UEs) simultaneously. In fully digital architectures, each base station (BS) antenna is equipped with a dedicated radio-frequency chain that includes complex, power-hungry analog-to-digital/digital-to-analog converters (ADCs/DACs) \cite{Xia17}. In this setting, the power consumed by each ADC/DAC scales linearly with the sampling rate and exponentially with the number of quantization bits \cite{Mo15,Cho16,Li17,Sax17}. Another limiting aspect is the volume of raw data exchanged between the remote radio head and the base-band unit, which scales linearly with both the sampling rate and the number of quantization bits \cite{Jac17}.

For these reasons, adopting low-resolution ADCs/DACs (e.g., with 1 to 4 quantization bits) can enable the implementation of fully digital massive MIMO arrays comprising hundreds (or even thousands) of antennas, which are necessary to operate in the mmWave and THz bands \cite{Jac17}. In this regard, 1-bit ADCs/DACs are particularly appealing due to their minimal power consumption and complexity \cite{Mo15,Atz21a}. Such a coarse quantization is suitable especially at very high frequencies, where high-order modulations may not be needed due to the huge bandwidths. There is a vast literature on massive MIMO with 1-bit ADCs/DACs. For instance, the capacity~of~the~\mbox{1-bit} quantized MIMO channel is characterized in \cite{Mo15}. The work in \cite{Cho16} proposes an efficient iterative method for near maximum likelihood detection (MLD) with 1-bit ADCs. The channel estimation and the uplink achievable rate with 1-bit ADCs are studied in \cite{Li17}. The spectral efficiency of single-carrier and orthogonal frequency-division multiplexing uplink systems with 1-bit ADCs is analyzed in \cite{Mol17}. Some of the results derived in \cite{Li17,Mol17} for 1-bit ADCs are extended to the multi-bit case in \cite{Jac17}. The performance of downlink linear precoding with 1-bit DACs is studied in \cite{Sax17}. The benefits of oversampling in massive MIMO systems with 1-bit ADCs are investigated~in~\cite{Ucu18}.

In this paper, we broaden prior analytical studies on the uplink data detection in massive MIMO systems with 1-bit ADCs. The statistical properties of the estimated symbols have not been characterized by previous works. In this respect, it was observed in \cite{Jac17} that the estimated symbols resulting from transmit symbols with the same phase overlap at high signal-to-noise ratio (SNR), although this aspect has not been formally described in the literature. We fill this gap by deriving closed-form expressions of the expected value and the variance of the estimated symbols for the single-UE case when maximum ratio combining (MRC) is adopted at the BS. Furthermore, we analyze their asymptotic behavior at high SNR. Building on these results, we propose an enhanced MLD method that considerably reduces the symbol error rate (SER) by properly weighting each detection region with the corresponding variance. Numerical results are presented to evaluate the SER with respect to the number of BS antennas, the SNR, and the pilot length used during the channel estimation phase. Our analysis highlights a fundamental SNR trade-off, according to which operating at the right SNR significantly improves the data detection accuracy.

\textit{Notation.} $\A = (A_{m,n})$ specifies that $A_{m,n}$ is the $(m,n)$th entry of matrix $\A$; likewise, $\a = (a_{n})$ specifies that $a_{n}$ is the $n$th entry of vector $\a$. The notation $\{ \cdot \}$ is used to represent sets, whereas $\Re[\cdot]$ and $\Im[\cdot]$ denote the real part and imaginary part operators, respectively.

\section{System Model} \label{sec:SM}

Let us consider a BS with $M$ antennas serving $K$ single-antenna UEs in the uplink. Each BS antenna is connected to a pair of 1-bit ADCs for the in-phase and the quadrature components of the receive signal. We thus introduce the 1-bit quantization function $Q(\cdot) : \Compl^{A \times B} \to \setQ$, with
\begin{align}
Q(\C) \triangleq \sqrt{\frac{\rho K + 1}{2}} \Big( \sgn \big( \Re[\C] \big) + j \, \sgn \big( \Im[\C] \big) \Big)
\end{align}

\clearpage

$ $ \vspace{-2.5mm}

\noindent and where $\setQ \triangleq \sqrt{\frac{\rho K + 1}{2}} \{ \pm 1 \pm j \}^{A \times B}$ \cite{Jac17}. We use $\H \in \Compl^{M \times K}$ to denote the uplink channel matrix whose entries are assumed to be distributed independently as $\setC \setN (0,1)$ (as, e.g., in \cite{Jac17,Li17}); more involved channel models will be considered in our future work. Furthermore, each UE transmits with power $\rho$ and the additive white Gaussian noise (AWGN) at the BS has unit variance: hence, $\rho$ can be interpreted as the transmit~SNR.

Let $x_{k} \in \Compl$ be the transmit symbol of UE~$k$, with $\Exp \big[ |x_{k}|^{2} \big] = 1$ and $\x \triangleq (x_{k}) \in \Compl^{K \times 1}$. The receive signal at the BS at the input of the ADCs is given by
\begin{align}
\y \triangleq \sqrt{\rho} \H \x + \z \in \Compl^{M \times 1}
\end{align}
where
$\z \in \Compl^{M \times 1}$ is the AWGN term with entries distributed as $\setC \setN (0,1)$. Then, at the output of the ADCs, we have
\begin{align} \label{eq:r}
\r \triangleq Q(\y) \in \Compl^{M \times 1}.
\end{align}
At this stage, the BS obtains a soft estimate of $\x$ as
\begin{align} \label{eq:x_hat}
\hat{\x} \triangleq \V^{\herm} \r \in \Compl^{K \times 1}
\end{align}
where $\V \in \Compl^{M \times K}$ is the combining matrix. Finally, the data detection process maps each estimated symbol to one of the transmit symbols.

\vspace{0.5mm}

\section{Data Detection Analysis with MRC} \label{sec:PA}

In this section, we focus on characterizing the performance of the data detection with respect to the different parameters when 1-bit ADCs are adopted at each BS antenna. In doing so, we consider the MRC receiver with combining matrix given by $\V = \hat{\H}$, where $\hat{\H} \in \Compl^{M \times K}$ is the estimate of $\H$ acquired during the uplink pilot-aided channel estimation phase. Let $\P \triangleq (P_{u,k}) \in \Compl^{\tau \times K}$ denote the pilot matrix whose columns correspond to the pilots used by the UEs, with $\{ |P_{u,k}|^{2} = 1 \}_{u,k}$, and where $\tau$ is the pilot length: assuming $\tau \geq K$ and orthogonal pilots among the UEs, we have $\P^{\herm} \P = \tau \I_{K}$. The UEs simultaneously transmit their uplink pilots and the receive signal at the BS at the input of the ADCs is given by
\begin{align}
\Y_{\rmp} \triangleq \sqrt{\rho} \H \P^{\herm} + \Z_{\rmp} \in \Compl^{M \times \tau}
\end{align}
where $\Z_{\rmp} \in \Compl^{M \times \tau}$ is the AWGN term with entries distributed as $\setC \setN (0,1)$. Then, at the output of the ADCs, we have
\begin{align} \label{eq:R_p}
\R_{\rmp} & \triangleq Q(\Y_{\rmp}) \in \Compl^{M \times \tau}.
\end{align}
Let us define \vspace{-0.5mm}
\begin{align}
\Omega(w) \triangleq \frac{2}{\pi} \arcsin(w)
\end{align}
and assume that $\hat{\H}$ is obtained via the scaled least-squares (LS) estimator
\begin{align} \label{eq:H_hat}
\hat{\H} \triangleq \sqrt{\Upsilon} \R_{\rmp} \P \in \Compl^{M \times K}
\end{align}
where we have defined
\begin{align} \label{eq:Upsilon}
\Upsilon & \triangleq \frac{2}{\pi} \frac{\rho}{(\rho K + 1)^2} \frac{\tau^{2}}{(\tau + \Delta)^{2}}
\end{align}
with
\begin{align}
\nonumber \Delta & \triangleq \frac{1}{K} \sum_{k=1}^{K} \sum_{u \neq v} \bigg( \Re[P_{u,k}^{*} P_{v,k}] \Omega \bigg( \frac{\rho \sum_{i=1}^{K} \Re[P_{u,i} P_{v,i}^{*}]}{\rho K + 1} \bigg) \\
\label{eq:Delta} & \phantom{=} \ - \Im[P_{u,k}^{*} P_{v,k}] \Omega \bigg( \frac{\rho \sum_{i=1}^{K} \Im[P_{u,i} P_{v,i}^{*}]}{\rho K + 1} \bigg) \bigg).
\end{align}
Note that the scaling factor in \eqref{eq:Upsilon} is chosen to minimize the mean squared error of the channel estimation for the class of scaled LS estimator: this is discussed in \cite{Atz21a}, which presents a detailed analysis of the channel estimation with 1-bit ADCs. Therefore, from \eqref{eq:x_hat}, the estimated symbols are obtained as $\hat{\x} = \sqrt{\Upsilon} \P^{\herm} \R_{\rmp}^{\herm} \r$. We point out that, when the MRC receiver results from the quantized channel estimation, it cannot be perfectly aligned with the channel matrix and results in residual multi-UE interference even when $M \to \infty$.

In this paper, we focus on the single-UE case (i.e., $K = 1$) and characterize the statistical properties of the estimated symbols.\footnote{Note that, when $K=1$, the scaled LS estimator in \eqref{eq:H_hat} with the scaling factor chosen as in \eqref{eq:Upsilon} is equivalent to the state-of-the-art linear estimator proposed in \cite{Li17}. We refer to \cite{Atz21a} for more details.} Hence, in this preliminary analysis, we do not consider the aforementioned multi-UE interference, which can be included at the expense of more involved and less insightful expressions: this will be explored in our future work.

\subsection{Expected Value and Variance of the Estimated Symbols} \label{sec:PA_ev}

Let $x \in \setS$ denote the transmit symbol of the UE, where $\setS \triangleq \{s_{\ell} \in \Compl\}_{\ell=1}^{L}$ represents the set of $L$ transmit symbols. Moreover, let $\hat{s}_{\ell}$ be the estimated symbol resulting from transmit symbol $s_{\ell} \in \setS$. Lastly, we use $\p \triangleq (p_{u}) \in \Compl^{\tau \times 1}$ to denote the pilot used by the UE. To facilitate the data detection process at the BS, for each $s_{\ell} \in \setS$, we are interested in deriving the closed-form expression of the expected value of $\hat{s}_{\ell}$, denoted by $\El \triangleq \Exp [\hat{s}_{\ell}]$. \vspace{-1mm}

\begin{theorem} \label{thm:s_mean} \rm{
Assuming $K = 1$ and MRC, for each transmit symbol $s_{\ell} \in \setS$, the expected value of the resulting estimated symbol $\hat{s}_{\ell}$ is given by
\begin{align}
\nonumber \El & = \sqrt{\frac{2}{\pi} \rho} M \frac{\tau}{\tau + \Delta} \sum_{u=1}^{\tau} p_{u}^{*} \bigg( \Omega \bigg( \frac{\rho \Re[p_{u} s_{\ell}]}{\sqrt{(\rho + 1)(\rho |s_{\ell}|^{2} + 1)}} \bigg) \\
\label{eq:s_mean} & \phantom{=} \ + j \, \Omega \bigg( \frac{\rho \Im[p_{u} s_{\ell}]}{\sqrt{(\rho + 1)(\rho |s_{\ell}|^{2} + 1)}} \bigg) \bigg)
\end{align}
with $\Delta$ defined in \eqref{eq:Delta}, which can be simplified for $K = 1$ as
\begin{align}
\nonumber \Delta & = \sum_{u \neq v} \bigg( \Re[p_{u}^{*} p_{v}] \Omega \bigg( \frac{\rho \Re[p_{u} p_{v}^{*}]}{\rho + 1} \bigg) \\
\label{eq:delta} & \phantom{=} \ - \Im[p_{u}^{*} p_{v}] \Omega \bigg( \frac{\rho \Im[p_{u} p_{v}^{*}]}{\rho + 1} \bigg) \bigg).
\end{align}}
\end{theorem}

\begin{IEEEproof}
See \cite[App.~V]{Atz21a}.
\end{IEEEproof} \vspace{1mm}

\noindent The result of Theorem~\ref{thm:s_mean} can be used towards the efficient implementation of MLD. Specifically, each estimated symbol can be mapped to one of the expected values $\{\mathsf{E}_{\ell}\}_{\ell=1}^{L}$, which are derived as in \eqref{eq:s_mean} without any prior Monte Carlo computation, according to the minimum distance criterion. To further reduce the data detection complexity, one can construct the Voronoi tessellation based on $\{\mathsf{E}_{\ell}\}_{\ell=1}^{L}$ obtaining well-defined detection regions: this allows to avoid the computation of the distance between each estimated symbol and each $\El$. It is worth mentioning that, in the case of multi-UE transmission, the expression in \eqref{eq:s_mean} will be conditioned on the symbols transmitted by all the UEs.

Now, for each $s_{\ell} \in \setS$, we are interested in deriving the closed-form expression of the variance of $\hat{s}_{\ell}$, denoted by \mbox{$\Vl \triangleq \Var [\hat{s}_{\ell}]$}. \vspace{-1mm}

\begin{theorem} \label{thm:s_var} \rm{
Assuming $K = 1$ and MRC, for each transmit symbol $s_{\ell} \in \setS$, the variance of the resulting estimated symbol $\hat{s}_{\ell}$ is given by
\begin{align}
\label{eq:s_var} \Vl & = \frac{2}{\pi} \rho M \frac{\tau^{2}}{\tau + \Delta} - \frac{1}{M} |\El|^{2}
\end{align}
with $\El$ and $\Delta$ given in \eqref{eq:s_mean} and \eqref{eq:delta}, respectively.}
\end{theorem}

\begin{IEEEproof}
See \cite[App.~VI]{Atz21a}.
\end{IEEEproof} \vspace{1mm}

\begin{figure*}[t!]
\centering
\begin{subfigure}{0.325\textwidth}
\centering
\includegraphics[scale=1]{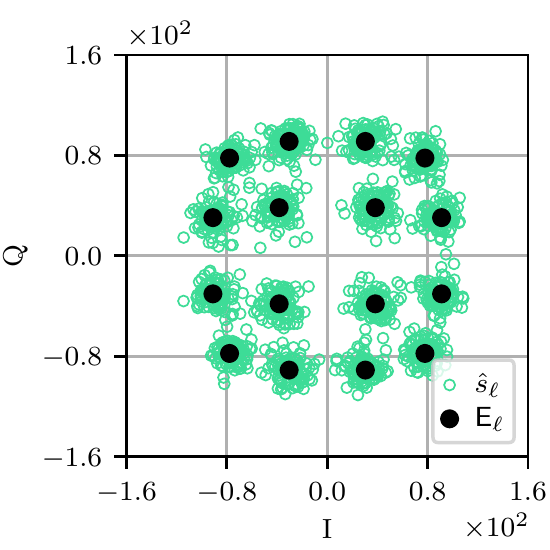}
\caption{$\rho = 0$~dB.}
\end{subfigure}
\begin{subfigure}{0.325\textwidth}
\centering
\includegraphics[scale=1]{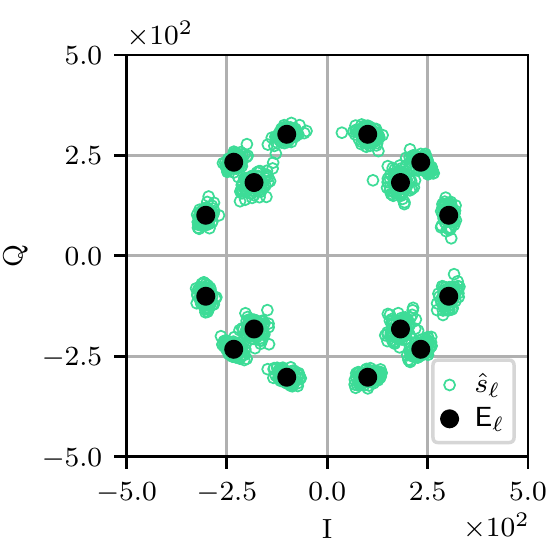}
\caption{$\rho = 10$~dB.}
\end{subfigure}
\begin{subfigure}{0.325\textwidth}
\centering
\includegraphics[scale=1]{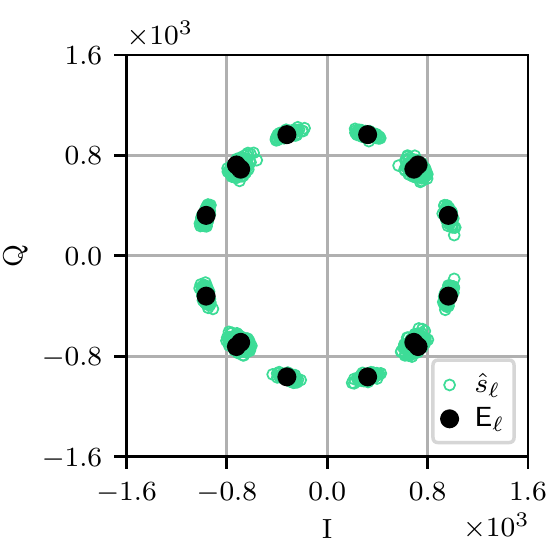}
\caption{$\rho = 20$~dB.}
\end{subfigure}
\caption{Estimated symbols with the MRC receiver, with 16-QAM transmit symbols, $M = 128$, and $\tau = 32$. The expected value of the estimated symbols is computed in closed form as in \eqref{eq:s_mean}.} \label{fig:dec_scatter} \vspace{-2mm}
\end{figure*}

\noindent The result of Theorem~\ref{thm:s_var} allows to quantify the absolute dispersion of the estimated symbols about their expected value, which arises from the 1-bit quantization applied to both the channel estimation (through the MRC receiver) and the uplink data transmission (see \eqref{eq:r}). This dispersion is not isotropic and assumes different shapes for different transmit symbols, as shown in Fig.~\ref{fig:dec_scatter} and in \cite{Abd21}. Furthermore, $\Vl$ diminishes~as $|s_{\ell}|$ increases due to the negative term on the right-hand side of \eqref{eq:s_var}, since the transmit symbols that lie further from the origin are less subject to noise. Let us now consider the normalized variance $\Vl/|\El|^{2}$, which quantifies the relative dispersion of $\hat{s}_{\ell}$ about its expected value. It is important to notice that, although $\Vl$ grows linearly with the number of BS antennas $M$, the normalized variance is inversely proportional to the latter.

The data detection process can be enhanced by taking into account the dispersion of the estimated symbols about their expected values. Specifically, in the context of MLD via Voronoi tessellation based on $\{\mathsf{E}_{\ell}\}_{\ell=1}^{L}$ described above, one can use the variance of the estimated symbols derived in \eqref{eq:s_var} to further refine the detection regions. In this setting, we adopt the approach of multiplicatively weighted Voronoi tessellation, where each detection region $\setR_{\ell}$ around $\El$ is constructed as
\begin{align} \label{eq:R_l}
\setR_{\ell} \triangleq \big\{ \xi \in \Compl : \omega_{\ell} |\xi - \El| \leq \omega_{i} |\xi - \mathsf{E}_{i}|, \forall i \neq \ell \big\}
\end{align}
where $\omega_{\ell} > 0$ is the weight corresponding to $\El$. In particular, one must choose each $\omega_{\ell}$ to be a decreasing function of $\Vl$ such that a higher variance of $\hat{s}_{\ell}$ corresponds to a smaller distance function around $\El$ and, consequently, gives rise to a larger $\setR_{\ell}$ (see, e.g., the choice in \eqref{eq:omega_l}).\footnote{Note that the case of equal weights corresponds to conventional MLD.} Remarkably, it is shown in Section~\ref{sec:NUM_voronoi_alpha} that this approach can greatly boost the performance of the data detection in terms of SER.

We now analyze the asymptotic behavior of the expected value and the variance of the estimated symbols at high SNR. \vspace{-1mm}

\begin{corollary} \label{cor:detection-lim_rho} \rm{
From Theorems~\ref{thm:s_mean}~and~\ref{thm:s_var}, in the limit of $\rho \to \infty$, we have
\begin{align}
\nonumber \lim_{\rho \to \infty} \frac{\El}{\sqrt{\rho}} & = \sqrt{\frac{2}{\pi}} M \frac{\tau}{\tau + \bar{\Delta}} \sum_{u=1}^{\tau} p_{u}^{*} \bigg( \Omega \bigg( \frac{\Re[p_{u} s_{\ell}]}{|s_{\ell}|} \bigg) \\
\label{eq:s_mean-lim_rho} & \phantom{=} \ + j \, \Omega \bigg( \frac{\Im[p_{u} s_{\ell}]}{|s_{\ell}|} \bigg) \bigg)
\end{align}
and
\begin{align}
\label{eq:s_var-lim_rho} \lim_{\rho \to \infty} \frac{\Vl}{\rho} & = \frac{2}{\pi} M \frac{\tau^{2}}{\tau + \bar{\Delta}} - \frac{1}{M} \lim_{\rho \to \infty} \frac{|\El|^{2}}{\rho}
\end{align}

\noindent where we have defined
\begin{align}
\bar{\Delta} & = \sum_{u \neq v} \Big( \Re[p_{u}^{*} p_{v}] \Omega \big( \Re[p_{u} p_{v}^{*}] \big) - \Im[p_{u}^{*} p_{v}] \Omega \big( \Im[p_{u} p_{v}^{*}] \big) \Big).
\end{align}}
\end{corollary}

\noindent The result of Corollary~\ref{cor:detection-lim_rho} formalizes a behavior of the estimated symbols that was observed in \cite{Jac17}. From \eqref{eq:s_mean-lim_rho}, at high SNR, all the estimated symbols lie on a circle around the origin and the information carried by the amplitude of the transmit symbols is entirely suppressed by the 1-bit quantization. Therefore, the estimated symbols resulting from transmit symbols with the same phase become indistinguishable in terms of their expected value, which depends only on $\Re[s_{\ell}]/|s_{\ell}|$ and $\Im[s_{\ell}]/|s_{\ell}|$. For example, if $\setS$ corresponds to the 16-QAM constellation (as considered in Section~\ref{sec:NUM}), the inner estimated symbols become indistinguishable from the outer estimated symbols with the same phase. Moreover, according to \eqref{eq:s_var-lim_rho}, these estimated symbols become identical also in terms of variance. In view of these aspects, the system performance cannot be enhanced simply by minimizing the normalized variance of the estimated symbols. On the one hand, such a variance roughly decreases with the transmit SNR; on the other hand, the overlap between different symbols after the estimation increases with the transmit SNR. This determines a clear SNR trade-off, according to which operating at the right SNR enhances the data detection accuracy.

\section{Numerical Results} \label{sec:NUM}

In this section, we evaluate the performance of the data detection with 1-bit ADCs with respect to the different parameters using the analytical results presented in Section~\ref{sec:PA_ev}. We assume that, during the uplink pilot-aided channel estimation phase, the second column of the $\tau$-dimensional discrete Fourier transform matrix is used as pilot, i.e., $\d_{2} \triangleq [1, e^{-j \, \frac{2 \pi}{\tau}}, e^{-j \, 2 \frac{2 \pi}{\tau}}, \ldots, e^{-j \, (\tau-1) \frac{2 \pi}{\tau}}]^{\tran} \in \Compl^{\tau \times 1}$, which represents the best possible pilot choice (see \cite[App.~I]{Atz21a} for more details). In addition, we assume the same transmit SNR for the two phases of channel estimation and uplink data transmission. Lastly, although our analytical framework is valid for any choice of the set of transmit symbols $\setS$, we analyze the scenario where $\setS$ corresponds to the 16-QAM constellation, i.e., $\setS = \frac{1}{\sqrt{10}} \big\{ \pm 1 \pm j, \pm 1 \pm j \, 3, \pm 3 \pm j, \pm 3 \pm j \, 3 \big\}$.\footnote{Note that the symbols are normalized such that $\frac{1}{L} \sum_{\ell=1}^{L} |s_{\ell}|^{2} = 1$.}

\begin{figure}[t!]
\centering
\includegraphics[scale=1]{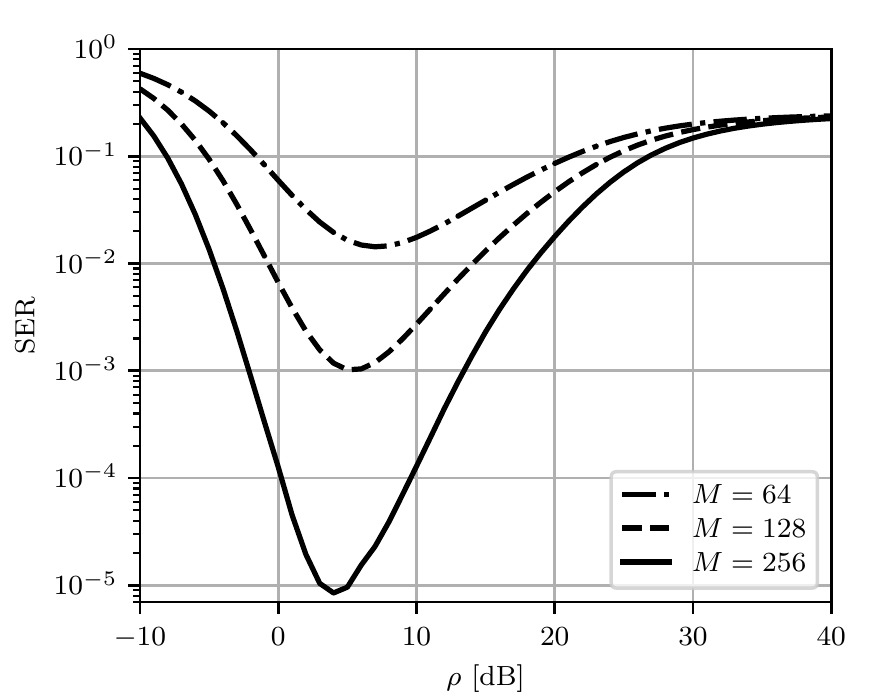}
\caption{SER against the transmit SNR, with 16-QAM transmit symbols, $M \in \{64, 128, 256\}$, and $\tau=32$.} \label{fig:SER_VS_rho} \vspace{-1mm}
\end{figure}

Fig.~\ref{fig:dec_scatter} illustrates the estimated symbols for different values of the transmit SNR $\rho$, with $M = 128$ and $\tau = 32$; each 16-QAM symbol is transmitted over $10^{2}$ independent channel realizations. The expected value of the estimated symbols is computed as in Theorem~\ref{thm:s_mean} and clearly matches the corresponding sample average. Here, we observe two fundamental and conflicting trends that constitute the SNR trade-off described in Section~\ref{sec:PA_ev}. First, the normalized variance of the estimated symbols decreases with the transmit SNR. Second, the estimated symbols resulting from the transmit symbols with the same phase, i.e., $\pm \frac{1}{\sqrt{10}} (1 \pm j)$ and $\pm \frac{1}{\sqrt{10}} (3 \pm j \, 3)$, get closer as the transmit SNR increases from $\rho = 0$~dB to $\rho = 10$~dB and almost fully overlap at $\rho = 20$~dB. This behavior was observed in \cite{Jac17} and is formalized in Corollary~\ref{cor:detection-lim_rho}, according to which such estimated symbols become identical at high SNR and the difference in amplitude between symbols cannot be recovered. For the 16-QAM, this produces a SER of $0.25$ since there are four pairs of indistinguishable estimated symbols (see also Fig.~\ref{fig:SER_VS_rho}).

\begin{figure}[t!]
\centering
\includegraphics[scale=1]{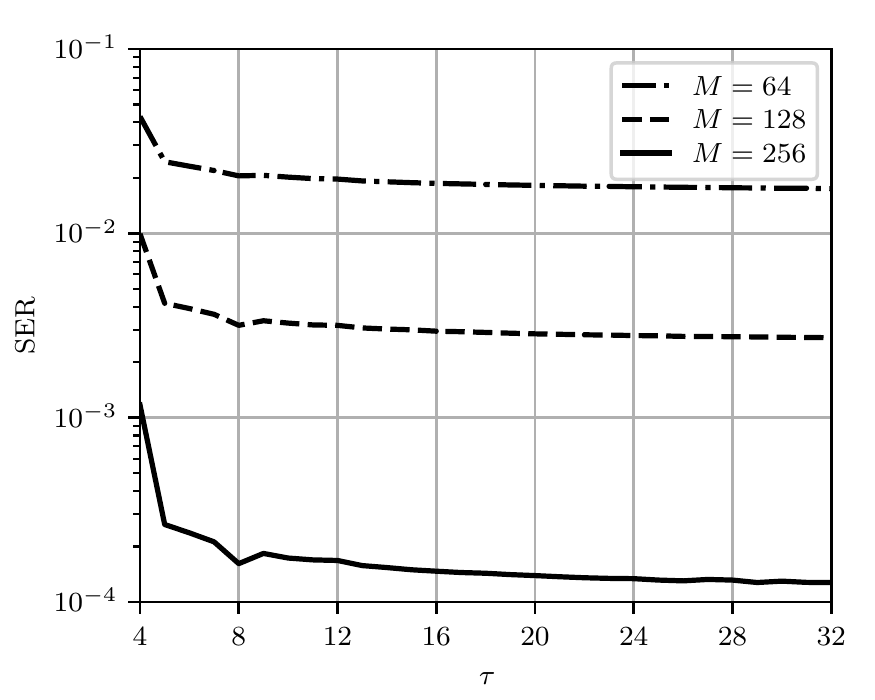}
\caption{SER against the pilot length, with 16-QAM transmit symbols, $M \in \{64, 128, 256\}$, and $\rho=10$~dB.} \label{fig:SER_VS_tau} \vspace{-1mm}
\end{figure}

We now examine the combined effect of the channel estimation and the data detection with 1-bit ADCs on the system performance in terms of SER, which is computed numerically via Monte Carlo simulations with $10^{6}$ independent channel realizations. The symbols are decoded via MLD aided by the result of Theorem~\ref{thm:s_mean}. Furthermore, different numbers of BS antennas are considered, i.e., $M \in \{64, 128, 256\}$. Fig.~\ref{fig:SER_VS_rho} plots the SER against the transmit SNR $\rho$, with $\tau = 32$, showing a clear SNR trade-off. In particular, the SER reduces until it attains its minimum (which occurs at about $\rho = 4$~dB for $M = 256$) before increasing again and reaching asymptotically the value of $0.25$. In fact, as discussed above for Fig.~\ref{fig:dec_scatter}, the inner estimated symbols of the 16-QAM constellation become indistinguishable from the outer estimated symbols with the same phase at high SNR. Fig.~\ref{fig:SER_VS_tau} depicts the SER against the pilot length $\tau$, with $\rho = 10$~dB, showing the impact of the channel estimation accuracy in the computation of the MRC receiver. For instance, for $M = 256$, the SER is decreased by a factor of $5$ when the pilot length grows from $\tau = 4$ to $\tau = 8$. We refer to \cite{Atz21a} for a thorough analysis of the channel estimation with 1-bit ADCs. In both Fig.~\ref{fig:SER_VS_rho}~and~\ref{fig:SER_VS_tau}, we observe that increasing the size of the antenna array at the BS is always beneficial. For example, in Fig.~\ref{fig:SER_VS_rho}, the SER is decreased by two orders of magnitude at the optimal transmit SNR when the number of BS antennas grows from $M = 128$ to $M = 256$. Indeed, the higher granularity in the antenna domain allows to sum the contribution of a larger number of independent channel entries.

\subsection{Enhanced Maximum Likelihood Detection} \label{sec:NUM_voronoi_alpha}

\begin{figure}[t!]
\centering
\begin{subfigure}{0.49\textwidth}
\centering
\includegraphics[scale=1]{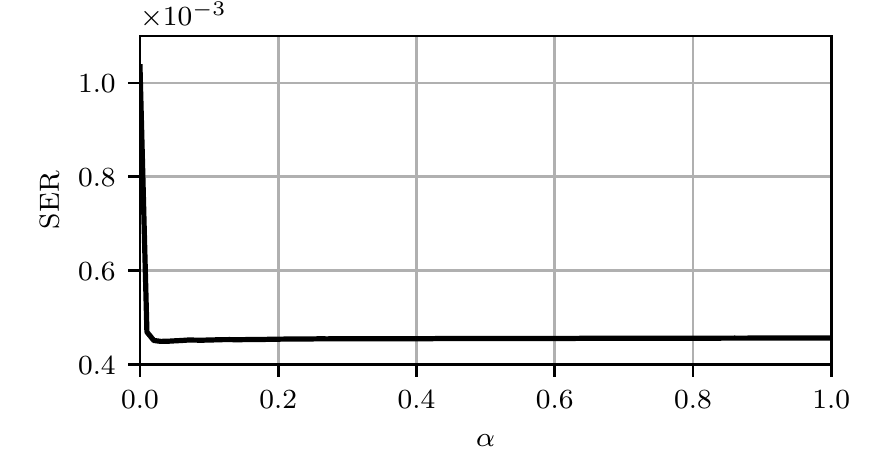}
\caption{SER against $\alpha$.}
\end{subfigure}
\begin{subfigure}{0.49\textwidth}
\centering
\includegraphics[scale=1]{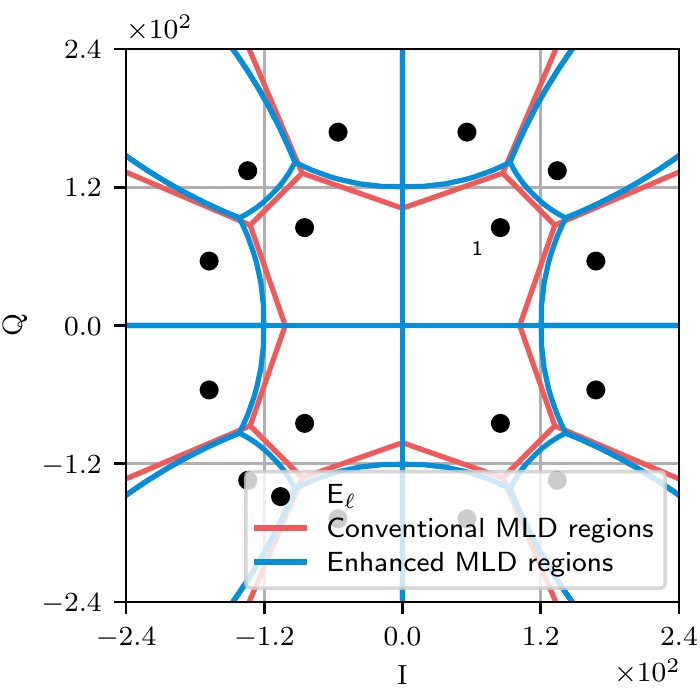}
\caption{Detection regions for conventional and enhanced MLD corresponding to $\alpha = 0$ and $\alpha = 1$, respectively.}
\end{subfigure}
\caption{Enhanced MLD with weights chosen as in \eqref{eq:omega_l}, with $M=128$, $\rho = 5$~dB, and $\tau=32$.} \label{fig:voronoi_alpha} \vspace{-1mm}
\end{figure}

The SER results presented so far have been obtained with conventional MLD, whereby each estimated symbol is mapped to one of the expected values $\{\mathsf{E}_{\ell}\}_{\ell=1}^{L}$ according to the minimum distance criterion. Such a data detection process can be enhanced by taking into account the dispersion of the estimated symbols about their expected values, i.e., by assigning larger detection regions to the estimated symbols with higher variance. Hence, we now construct the detection regions according to a multiplicatively weighted Voronoi tessellation (see \eqref{eq:R_l}) with the following heuristic choice of the weights:
\begin{align} \label{eq:omega_l}
\omega_{\ell} = \frac{1}{1+\alpha(\Vl-1)}, \qquad \ell = 1, \ldots, L
\end{align}
with $\alpha \in [0,1]$. This choice allows to strike a balance between conventional MLD (i.e., $\omega_{\ell} = 1$ for $\alpha = 0$) and enhanced MLD with weights inversely proportional to the variance of the estimated symbols (e.g., $\omega_{\ell} = 1/\Vl$ for $\alpha = 1$).

Fig.~\ref{fig:voronoi_alpha}(a) plots the SER against $\alpha$, with $M=128$, $\rho = 5$~dB, and $\tau=32$, showing that using even slightly weighted detection regions can reduce the SER by a factor of $2$. Fig.~\ref{fig:voronoi_alpha}(b) illustrates the detection regions corresponding to the cases of $\alpha = 0$ and $\alpha = 1$. It is straightforward to observe that the detection regions corresponding to the inner estimated symbols of the 16-QAM constellation (with higher variance) are enlarged at the expense of the ones corresponding to the outer estimated symbols (with lower variance). For instance, the detection threshold between the estimated symbols corresponding to $\frac{1}{\sqrt{10}} (1 + j)$ and $\frac{1}{\sqrt{10}} (3 + j \, 3)$ is shifted outwards to accommodate the larger dispersion of the former (cf. Fig.~\ref{fig:dec_scatter}). Indeed, this simple approach can greatly boost the performance of the data detection in terms of SER.

\section{Conclusions} \label{sec:CONCL}

This paper focuses on the uplink data detection analysis of massive MIMO systems with 1-bit ADCs. We characterize the expected value and the variance of the estimated symbols when MRC is adopted at the BS along with their asymptotic behavior at high SNR. Building on these results, we propose an enhanced MLD method that is able to greatly reduce the SER by taking into account the dispersion of the estimated symbols about their expected values. The proposed analysis provides important practical insights into the design and the implementation of 1-bit quantized systems: in particular, it highlights a fundamental SNR trade-off, according to which operating at the right SNR considerably improves the data detection accuracy. Future work will consider extensions to the multi-UE case and the optimal design of the set of transmit symbols capitalizing on our analytical framework.

\bibliographystyle{IEEEtran}
\bibliography{IEEEabrv,refs}

\end{document}